\documentclass[aps,prd,preprintnumbers,nofootinbib,floatfix,preprint]{revtex4}
\usepackage{xcolor}
\usepackage{graphicx}
\usepackage{wrapfig}
\usepackage{amsmath}
\usepackage{amsfonts}
\usepackage{amssymb}
\usepackage{multirow}
\usepackage{slashed}
\usepackage{physics}
\usepackage{float}
\usepackage{dcolumn}
\usepackage{color,soul}
\usepackage{comment}
\usepackage[colorlinks=true, linkcolor=blue, citecolor=blue, urlcolor=blue]{hyperref}
\usepackage[textheight=9in, textwidth=6.5in, letterpaper]{geometry}

\newcommand\befs{\begin{figure*}}
\newcommand\eefs[1]{\label{fig:#1}\end{figure*}}
\newcommand\bef{\begin{figure}}
\newcommand\eef[1]{\vskip -0.125cm \label{fig:#1}\end{figure}}
\newcommand\beq{\begin{equation}}
\newcommand\eeq[1]{\label{#1}\end{equation}}
\newcommand\beqa{\begin{eqnarray}}
\newcommand\eeqa[1]{\label{#1}\end{eqnarray}}
\newcommand\bet{\begin{table}}
\newcommand\eet[1]{\label{tb:#1}\end{table}}
\newcommand\bets{\begin{table*}}
\newcommand\eets[1]{\label{tb:#1}\end{table*}}
\newcommand{\be}{\begin{equation}}
\newcommand{\ee}{\end{equation}}
\newcommand{\bea}{\begin{eqnarray}}
\newcommand{\eea}{\end{eqnarray}}
\newcommand\fgn[1]{Fig.\ \ref{fig:#1}}
\newcommand\eqn[1]{Eq.\ (\ref{#1})}
\newcommand\scn[1]{Section \ref{sec:#1}}

\begin{document}

\date{\today}

\title{Approaching the conformal WZW  behavior in the infrared limit of two-dimensional massless QCD: a lattice study
}

\author{Nikhil\ \surname{Karthik}}
\email{nkarthik.work@gmail.com}
\affiliation{American Physical Society, Hauppauge, New York 11788\\
Department of Physics, Florida International University, Miami, FL 33199}
\author{Rajamani\ \surname{Narayanan}}
\email{rajamani.narayanan@fiu.edu}
\affiliation{Department of Physics, Florida International University, Miami, FL 33199}
\author{Sruthi A.\ \surname{Narayanan}}
\email{sruthi81294@gmail.com}
\affiliation{ Perimeter Institute for Theoretical Physics, Waterloo, ON N2L 2Y5, Canada }

\begin{abstract}
Two-dimensional QCD with $N_c$ colors and $N_f$ flavors of massless fermions in the fundamental representation is expected to exhibit conformal behavior in the infrared governed by a $u(N_f)$ WZW model with level $N_c$. Using numerical analysis within the lattice formalism with exactly massless overlap fermions, we show the emergence of such behavior in the infrared limit. Both the continuum extrapolated low-lying eigenvalues of the massless Dirac operator and  the propagator of scalar mesons exhibit a flow from the ultraviolet to the infrared.  We find that the amplitude of the conserved current correlator 
remains invariant under the flow, while the amplitude of the scalar correlator approaches $N_f$-independent values in the infrared.

\end{abstract}

\maketitle

\section{Introduction}\label{sec:intro}

Two-dimensional QCD serves as an interesting toy model to study emergent conformal behavior in the infrared. As long as the number of colors, $N_c> 1$, is finite, the $su(N_c)$ gauge theory with $N_f$ flavors of massless fermions in the fundamental representation has a conformal sector in the infrared limit~\cite{Delmastro:2021otj}.
The identification of which two-dimensional conformal field theory describes the massless sector of the theory is somewhat of a conjecture. An ab initio numerical study of the infrared limit of QCD and verification of the existing expectation that the long-distance behavior of two-dimensional QCD is described by an appropriate Wess–Zumino–Witten (WZW) model is the aim of this paper.

Two-dimensional QCD is usually studied in the Hamiltonian formalism, and light-cone gauge serves as a convenient choice to work in. With this gauge choice, in addition to the gauge field being constrained, the fermion of one chirality (e.g. the right-chiral $\psi_R$) is also constrained, leaving the dynamics to be governed by a left-chiral fermion $\psi_L$. It is additionally advantageous to work on the light-cone as the momentum takes on only positive values, which become discrete if the direction perpendicular to the propagation direction is compact. This enables one to study the problem numerically via the method referred to as the Discrete Light-Cone Quantization (DLCQ)~\cite{Hornbostel:1988fb,Hornbostel:1988ne}. The flavor current constructed from the left-chiral fermion commutes with the interacting Hamiltonian, which fully describes the dynamics in the massless limit. Therefore, the states formed by repeatedly acting on the vacuum with flavor current operators are massless states. However, the flavor currents formed by the constrained right-chiral fermions do not commute with the Hamiltonian, and we do not have a proper description of the conformal behavior such as that given by a WZW model. 

Instead, if one considers the theory not on the light-front in order to take all degrees of freedom into account, one could conjecture~\cite{Delmastro:2021otj} that the infrared description is from a WZW model. The basis of this statement comes from an identification that the infrared limit, given by the energy scales $E\to 0$, is naively the same as the coupling $g^2\to\infty$ limit. In practice, one always has an upper cut-off either from a truncation of the Hilbert space to diagonalize the Hamiltonian associated with a timelike direction, or from the inverse lattice spacing, which can obscure the $g^2\to\infty$ limit. 

In this work, we impose the ultraviolet regulator using a lattice discretization and study the infrared limit carefully.
We will work within the Euclidean lattice formalism, where the flavor symmetry remains intact. Furthermore, we will use exact chiral fermions on the lattice even away from the continuum limit, which will result in the full flavor symmetry of the left- and right-chiral currents. 
Of particular interest to us in this paper will be the scaling dimension of the scalar mesons,
\be
M_{ij}(x) = \psi^\dagger_{iL}(x) \psi_{jR}(x) + \psi^\dagger_{iR}(x) \psi_{jL}(x)
\label{mesonop}
\ee
where the color indices that are summed over are suppressed and only the flavor indices, $i,j=1,\cdots,N_f$, are explicitly shown. 

We plan to study $su(N_c)$ gauge theories with exactly massless fermions using the lattice formalism in this paper. Since there are no topological zero modes, the one-point function will be exactly zero in any gauge field background, and we do not need to distinguish between $SU(N_f)$ scalars and the $U(1)$ scalar.  In this case, the expected scaling dimensions of $M_{ij}$ are that of a $u(N_f)$ WZW model with level $N_c$ and we cannot directly use the formula for the scaling dimension in~\cite{Delmastro:2022prj}. To that end, we provide a derivation of the formula for the scaling dimensions of the scalar mesons within the context of the WZW model in \scn{WZW}. This is followed by a description of the lattice formulation of two-dimensional QCD with exactly massless fermions in \scn{lattice}. The dimensionless length of the symmetric Euclidean torus, $\ell$, will be the only parameter in the continuum limit of the lattice formalism. We will use the low-lying eigenvalues of the massless lattice Dirac operator and  the meson correlator to show that the scaling dimension associated with the WZW algebra match the behavior as $\ell\to\infty$ in \scn{results}. In addition, we will also study the vector meson (current) correlator and show that it remains invariant under the flow from ultraviolet to infrared.

\section{WZW algebra}\label{sec:WZW}
The Wess-Zumino-Witten model~\cite{DiFrancesco:1997nk} of relevance to us is given by the action
\begin{equation}
S^{U(N_f)}_{N_c} = \frac{N_c}{16\pi}\int d^2x \mbox{Tr}(\partial^\mu g^{-1}\partial_\mu g)  + N_c\Gamma
\end{equation}
where  $g \in U(N_f)$  is identified as the affine-primary\footnote{The primary with respect to the affine current defined in the theory.} field in the CFT and $\Gamma$ is the Wess-Zumino term given by
\begin{equation}
\Gamma = -\frac{i}{24\pi}\int_{\mathcal{C}} d^3 y \epsilon_{\alpha\beta\gamma}\mbox{Tr}\left(\tilde{g}^{-1}\partial^\alpha \tilde{g} \tilde{g}^{-1}\partial^\beta \tilde{g} \tilde{g}^{-1}\partial^\gamma \tilde{g}\right).
\end{equation}
This is understood as an integral over a three-dimensional manifold $\mathcal{C}$ where $\tilde{g}$ is the extension of the primary fields to this manifold. Note that the level of this WZW model is given by $\frac{N_c}{2}$. The infrared limit of the mesons, $M_{ij}$, are expected to be associated with $g$ so our aim, in this section, is to obtain an expression for the scaling dimension of $g$. It is possible to extract this from a formula in~\cite{DiFrancesco:1997nk} but we provide some details because the group is neither simple nor semisimple.

This WZW model admits conserved flavor currents $J(z)=J_L(z),\bar{J}(\bar{z})=J_R(\bar{z})$ (the left-chiral and right-chiral currents in \scn{intro}) which can be expressed in terms of $g$ as 
\begin{equation}
J_L(z) = -\frac{N_c}{2} \partial_z g g^{-1}, \ \ J_R(\bar{z}) = \frac{N_c}{2} g^{-1}\partial_{\bar{z}}g.
\end{equation}
The modes of these currents satisfy the affine Kac-Moody algebra given by
\be
\left[ J_{k_1f_1\{R,L\}}, J_{k_2f_2\{R,L\}}\right] = if_{f_1 f_2 f_3}J_{(k_1+k_2)f_3\{R,L\}} +  \frac{N_c k_1}{2}\delta(k_1+k_2).\label{affine}
\ee In what follows, we drop the dependence on $z,\bar{z}$ and just notate the chirality of the current. Most generally, one constructs a stress tensor from a bilinear of these currents as per the Sugawara construction
\begin{equation}
\mathcal{T} = \sum_a \mathcal{C}_a :J_L^a J_L^a:
\end{equation}
where $: \ :$ denotes normal ordering and $\mathcal{C}_a$ are a set of constants picked to normalize the stress tensor such that it has the usual operator product expansion. Mode expanding the Sugawara stress tensor, as per standard two-dimensional CFT, one obtains the following identification of the Virasoro generators
\begin{equation}
\mathcal{L}_n = \sum_a\mathcal{C}_a\left[\sum_{k<0}J_{kL}^a J_{n-k,L}^a + \sum_{k\geq 0}J_{n-k,L}^a J_{kL}^a\right] = \sum_a\mathcal{C}_a L_n^a
\end{equation}
where the sum has been divided in order to impose the normal ordering and $L_n^a$ is a notation that will be useful in what follows. 

We want to know the explicit expression for the Sugawara stress tensor when the $J^a_L$ are $u(N_f)$ currents. In this case, the stress tensor can be decomposed into two parts, one that has all the $su(N_f)$ information and a $u(1)$ piece.\footnote{Usually one says that when you have a semisimple Lie algebra, the Sugawara stress tensor is the sum of the stress tensors for each simple Lie algebra. However, $U(N)$ is not semisimple, so we cannot make that argument here.} This should be reminiscent of the fact that extending from $su(N_f)$ to $u(N_f)$ requires adding a generator (the identity) that has a different scaling factor than the other generators. Computing the commutators of these modes gives, most generally,
\begin{eqnarray}
\left[\mathcal{L}_n,\mathcal{L}_m\right] & = & (n-m)\left[(N_c+N_f)\mathcal{C}^2\sum_{b=1}^{N_f^2-1}L_n^b+N_c\mathcal{C}_0^2L_n^0\right]\cr
& + & \left[\frac{N_c}{2}(N_c+N_f)\mathcal{C}^2(N_f^2-1)+\frac{N_c^2}{2}\mathcal{C}_0^2\right]\frac{n(n^2-1)}{6}\delta_{n+m}.
\end{eqnarray}
Here $\mathcal{C}$ denotes the $su(N_f)$ normalization and $\mathcal{C}_0$ is the $u(1)$ normalization constant. Asserting that this has to be the usual Virasoro algebra, allows us to solve for the normalization constants and write the $u(N_f)$ Sugawara stress tensor as 
\begin{equation}
\mathcal{T}_{u(N_f)} = \frac{1}{N_c+N_f}\sum_{a=1}^{N_f^2-1}:J^a_LJ^a_L: + \frac{1}{N_c}:J^0_LJ^0_L:.
\end{equation}

Now that we have the stress tensor, we can find the dimension of the primary $g$. This relies on the fact that in addition to an affine primary, $g$ is also a Virasoro primary. In any two-dimensional CFT, the operator product expansion (OPE) of the stress tensor with a primary field $\phi$ is given by~\cite{DiFrancesco:1997nk}
\begin{equation}
T(z)\phi(w,\bar{w}) \sim \frac{h}{(z-w)^2}\phi(w,\bar{w})+\frac{1}{z-w}\partial_w\phi(w,\bar{w}).
\end{equation}
The conformal weight $h$ of the primary is given by the coefficient of the most singular term. Therefore, if we compute the most singular term of the OPE between $\mathcal{T}_{u(N_f)}(z)$ and the WZW primary $g(w,\bar{w})$, we can read off its weight. In a WZW model the OPE between the current and the primary field $g$ is given by~\cite{DiFrancesco:1997nk}
\begin{equation}
J^a(z)g(w,\bar{w})\sim -\frac{T^ag(w,\bar{w})}{z-w}.
\end{equation}
Combining this with the definition of the $u(N_f)$ stress tensor and using identities of the sums of products of the generators $T^a$, one finds the weight of the primary to be\footnote{We note that Jaume Gomis has independently verified this formula via a private communication.}
\begin{equation}
h_g = \frac{N_f^2-1}{2N_f(N_c+N_f)} + \frac{1}{2N_cN_f}= \frac{N_cN_f +1}{2N_c (N_c+N_f)}.
\end{equation}
Additionally, one can read off the central charge as $c = \frac{N_f(N_cN_f+1)}{N_c+N_f}$ which was verified in~\cite{Naculich:2007nc} via alternative methods. 

In the following analysis, we concern ourselves with the scaling dimension of the scalar mesons given in~\eqref{mesonop}. This operator can be written in terms of the WZW primary as $M_{ij}(x) = g_{ij}+g_{ij}^\dagger$. One expects the conformal two-point function of two mesons to have the behavior $\frac{1}{|x-y|^{2\Delta}}$ where $\Delta=2h_g$ is the conformal dimension. In terms of $\Delta$, the mass anomalous dimension is given by  
\begin{equation}\label{weight}
\gamma_m = 1-\Delta = \frac{N_c^2-1}{N_c(N_c+N_f)}.
\end{equation}

Our numerical results are for $(N_c,N_f)=(2,1),(2,2),(3,1),(3,2)$ and the specific values of $\Delta$ are
$\frac{1}{2},\frac{5}{8},\frac{1}{3},\frac{7}{15}$ respectively.

\section{Lattice formalism}\label{sec:lattice}
\subsection{Lattice action and Monte Carlo simulation}
We study the theory on a finite two-dimensional Euclidean periodic box
of physical extent $\ell\times\ell$, where $\ell$ is the dimensionless length measured in units of the 't Hooft coupling. Let the lattice be denoted as $L\times L$ where we assume $L$ to be even. We will obtain the continuum limit at a fixed value of $\ell$ by extrapolating the results obtained at a few finite values of $L$ in the limit as $L\to\infty$. The lattice path-integral for the two-dimensional QCD 
coupled to $N_f$ flavors of massless fermions is
\beq
Z = \left(\prod_{x,\mu}\int dU_\mu(x)\right) \det\left(D_o\right)^{N_f} e^{S_g},
\eeq{pathint}
where $D_o$ is the overlap Dirac operator. $S_g$ is the 
plaquette gauge action which is written in terms of gauge-links $U_\mu(x)$ below
\be
S_g=\frac{N_c L^2 }{\ell^2}\sum_x \left[2N_c-P(x) -P^*(x)\right],\qquad
P(x) = \Tr \left[ 
U_1(x)U_2(x+\hat 1)U^\dagger_1(x+\hat 2)U^\dagger_2(x)\right],
\ee
where the trace is over color. To see the connection with the standard way to introduce lattice coupling in the action, note that
$\ell = \sqrt{\lambda} \ell_{\rm ph}$ where $\lambda$ is the dimensionful 't Hooft coupling and $\ell_{\rm ph}$ is the dimensionful length. Then $\frac{N_c L^2}{\ell^2} =\frac{1}{(ga)^2}$ where $\lambda=g^2 N_c$ and $a=\ell_{\rm ph}/L$ is the lattice spacing.
The overlap-Dirac~\cite{Narayanan:1994gw,Neuberger:1997fp,Edwards:1998wx} operator $D_o$ can be written as 
\be
D_o = \frac{1+V}{2},
\ee
in terms of a unitary operator $V = \sigma_3 \epsilon(H_w)$ written in terms of the sign-function $\epsilon$ and the Hermitian Wilson-Dirac operator $H_w = \sigma_3 D_w$. We approximated the sign function using a 21-pole Zolotarev rational approximation. In two dimensions, the Wilson-Dirac operator that enters the construction of the overlap operator can be written as
\be
D_w=\left(2-M\right) - \frac{1}{2}\left[ \left(1-\sigma_1\right)T_1+ \left(1+\sigma_1\right)T_1^\dagger+
 \left(1-\sigma_2\right)T_2+ \left(1+\sigma_2\right)T_2^\dagger\right],
 \ee
where the Wilson mass $M\in (0,2)$. We set $M=1$ in this study. In the above equation, $\sigma_\mu$ are the Pauli matrices, and $T_1$ and $T_2$ are the gauge covariant parallel transporters,
\be
(T_\mu \phi)(x) = U_\mu(x) \phi(x+\hat\mu),\qquad (T_\mu^\dagger \phi)(x) = U_\mu^\dagger(x-\hat\mu)\phi(x-\hat\mu).
\ee
In two dimensions, the lack of topological sectors in the 
$su(N_c)$ theory implies that there are no topological zero modes
for the overlap Dirac operator. This lets us make the following simplification: noting that 
\be
H_o = \sigma_3 D_o = \frac{ \sigma_3 + \epsilon(H_w)}{2},\qquad \left[ H_o^2 , \sigma_3\right]=0
\ee
we can write
\be
\det D_o = \det H_o = \det_+ H_o^2 = \det_- H_o^2
\ee
where $\pm$ restricts the determinant to one chiral sector.

We used the standard HMC algorithm~\cite{Duane:1987de} to sample gauge configuration weighted according to \eqn{pathint}. Whereas the implementation of the HMC trajectories via the leap frog method is standard, we note the following interesting feature. The fermionic action using $N_f$ copies of pseudofermions $\phi_{i+}$ is given by
\be
S_f(\phi)=\sum_{i=1}^{N_f}\phi_{i+}^\dagger \frac{1}{H_o^2} \phi_{i+},\qquad \sigma_3 \phi_{i+}=\phi_{i+}.
\ee
In the above equation, by restricting the action to a single chiral sector, we are able to write the action in terms of a positive definite operator for each flavor, which thereby enables us to set $N_f$ to any value. The details about the algorithm are the same as the one employed in the study of three-dimensional QED in~\cite{Karthik:2016ppr}.

\subsection{Lattice observables to probe the conformal infrared behavior}
We study the scaling dimension of the scalar operator using the two-point function, as well as by probing the finite-size scaling of the 
low-lying eigenvalues of the overlap Dirac operator. In addition, we will also study the correlator of the vector meson in order to show that it does not acquire an anomalous dimension. Both the scalar and vector meson correlators can be written in terms of the anti-Hermitian massless overlap fermion propagator,
\be
G_o = \frac{1-V}{1+V}.
\ee
We first we notice that $\{G_o,\sigma_3\} =0$, implying that the form of the 
overlap propagator is
\be
G_o = \begin{pmatrix} 0 & G_L \cr -G_L^\dagger & 0 \end{pmatrix},
\ee
where $G_L$ is the propagator of the left-chiral fermion and $-G_L^\dagger$ is the propagator of the right-chiral fermion. 

Observables with better sensitivity to the infrared conformality are
the microscopic low-lying eigenvalues of the Dirac operator. For the 
overlap operator, they are the eigenvalues of the inverse of the overlap 
propagator $A=G_o^{-1}$, given by
\be
A v_j = i \Lambda_j v_j,
\ee
determined over a fixed gauge field background. Since eigenvalues come in 
positive-negative pairs, we can take $\Lambda_j>0$ in the above equation, 
and the eigenvalues are sorted as $\Lambda_{j+1} > \Lambda_j$ 
for $j=1,\cdots,N_c L^2$. We will only consider the first few values of 
$j=1,2,3,4$ in this study.  

We measured $\Lambda_j$ in the 
sampled gauge configurations, and determined the ensemble averages 
$\langle \Lambda_i\rangle$.  The continuum limits $\lambda_i (\ell)$ of the 
eigenvalues are given by
\be
\lambda_i (\ell) \ell \equiv \lim_{L\to\infty} \langle \Lambda_i \rangle L.
\ee
The dependence of the scalar susceptibility (i.e., scalar correlator 
integrated over the entire box), gives the corresponding 
expectation for the finite-size scaling of $\lambda_i(\ell)$. Namely,
again anticipating a free-theory-like short-distance behavior and a 
nontrivial conformal behavior in the infrared given by~\eqref{weight},
we expect
\be
\lambda_i(\ell) \ell \sim \ell^0 \ \ {\rm as} \quad \ell\to 0,\qquad{\rm and}\qquad 
\lambda_i(\ell) \ell \sim \frac{1}{\ell^{\gamma_m}},\quad \gamma_m = \frac{N_c^2-1}{N_c(N_f+N_c)} \ \ {\rm as}\quad \ell\to\infty.
\ee

Next, we turn our attention to meson correlators.
At a fixed value of $\ell$, the lattice scalar and vector meson correlators will be periodic functions that depend on the lattice separation, $X$. Upon taking the continuum limit, the correlators will naturally become a function of $\xi=\frac{x}{\ell}=\lim_{L\to\infty} \frac{X}{L}$. Usually, one studies the correlator as a function of $x$ as $\ell\to\infty$, which amounts to the behavior as $\xi\to 0$ at fixed $\xi\ell=x$.
In this limit, one recovers the correlators of operators $O$ on an infinite plane, which will behave as $C_O/x^{2\Delta_O}$ in the UV and IR limits, with the 
appropriate UV and IR values of the scaling dimensions $\Delta_O$. The coefficients of the power-law dependence, $C^{\rm UV}_O$ and $C^{\rm IR}_O$, in the two limits, are referred to as the UV and IR amplitudes. 
We can instead study the behavior of the correlators as a function of $\ell$ at a fixed $\xi > 0$ (see for example~\cite{Karthik:2016ppr}).
On a periodic torus, the correlators will still behave as power-law in operator separation $\xi\ell$ in the large and small $\ell$ limits, and the power-law  coefficients generalize into functions of $\xi$ and $\ell$ that approach their UV and IR limits, $C^{\rm UV}_O(\xi)$ and $C^{\rm IR}_O(\xi)$. The amplitudes of the power law on an infinite plane are then recovered from $C^{\rm UV, IR}_O(\xi)$ in the $\xi\to 0$ limit. 
In this work, we focus on these coefficients at a finite $\xi$ 
and study their flow from UV to IR to avoid additional, 
difficult $\xi\to0$ extrapolation.
Below, we apply this general discussion to the specific cases of the scalar meson and vector meson (conserved current) correlators.

In terms of $G_L$, we write the continuum pseudoscalar and the left-chiral and the right-chiral current correlators as
\bea
\ell^2 G_s(\ell,\xi) &=& \frac{1}{G_f(\xi)} \lim_{L\to\infty} L^2\left\langle\Tr \left[ G_L\left(0,L\xi\right) G_L^\dagger\left(0,L\xi\right) + G_L\left(L\xi,0\right) G_L^\dagger\left(L\xi,0\right)\right]\right\rangle,\cr
\ell^2 G_v(\ell,\xi) &=& \frac{1}{G_f(\xi)} \lim_{L\to\infty} L^2\left\langle\Tr \left[ G_L\left(0,L\xi\right) G_L\left(0,L\xi\right) + G^\dagger_L\left(L\xi,0\right) G_L^\dagger\left(L\xi,0\right)\right]\right\rangle.\label{correlators}
\eea
The trace is only over the color indices.
The factor 
$\ell^2$ multiplying the correlator on the left accounts for the na\"ive scaling dimension from units of lattice spacing to the 
units of the coupling. The free correlator, $G_f(\xi)$  is specific to the overlap-Dirac operator  and is defined as
\be
G_f(\xi) = \lim_{L\to\infty} L^2 \left[ G_{fL}\left(0,L\xi\right) G_{fL}^\dagger\left(0,L\xi\right) + G_{fL}\left(L\xi,0\right) G_{fL}^\dagger\left(L\xi,0\right)\right]
\ee
where $G_{fL}$ is the free fermion propagator. The free correlator will be proportional to $\frac{1}{\xi^2}$ when $\xi\to 0$.
Note that we have divided the correlators by $G_f(\xi)$ in \eqn{correlators} to take care of continuum-like normalization, and to avoid 
trivial factors of $\frac{1}{\xi^2}$.

We can define amplitudes at finite $\ell$ and $\xi$ from correlators by scaling with appropriate powers of $x = \xi\ell$. The scalar correlator is anomalous, and we will need to use two different scalings in the ultraviolet and infrared to define appropriate amplitudes, namely,
\be
C^{\rm UV}_s(\ell,\xi)=\ell^2 G_s(\ell,\xi)  \qquad C^{\rm IR}_s(\ell,\xi) =\left(\xi\ell\right)^{-2\gamma_m} \ell^2 G_s(\ell,\xi). 
\label{csuvir}
\ee
Then, $C^{\rm UV}_s(0,\xi)$ and $C^{\rm IR}_s(\infty,\xi)$ will give us the amplitudes in the ultraviolet and infrared.
In contrast, the vector meson correlator has the same scaling dimension in the ultraviolet and infrared, and we define its amplitude as
\be
C_v(\ell,\xi)=\ell^2 G_v(\ell,\xi).
\label{cvuvir}
\ee
Then, $C_v(0,\xi)$ and $C_v(\infty,\xi)$ will give us the amplitudes in the ultraviolet and infrared.
In this work, we will study the behavior of the RG flow of 
$C^{\rm UV}_s(\ell,\xi)$, $C^{\rm IR}_s(\ell,\xi)$ and $C_v(\ell,\xi)$ at a fixed value of $\xi=1/4$.

\section{Results}\label{sec:results}

\begin {figure}
\includegraphics[scale=0.34]{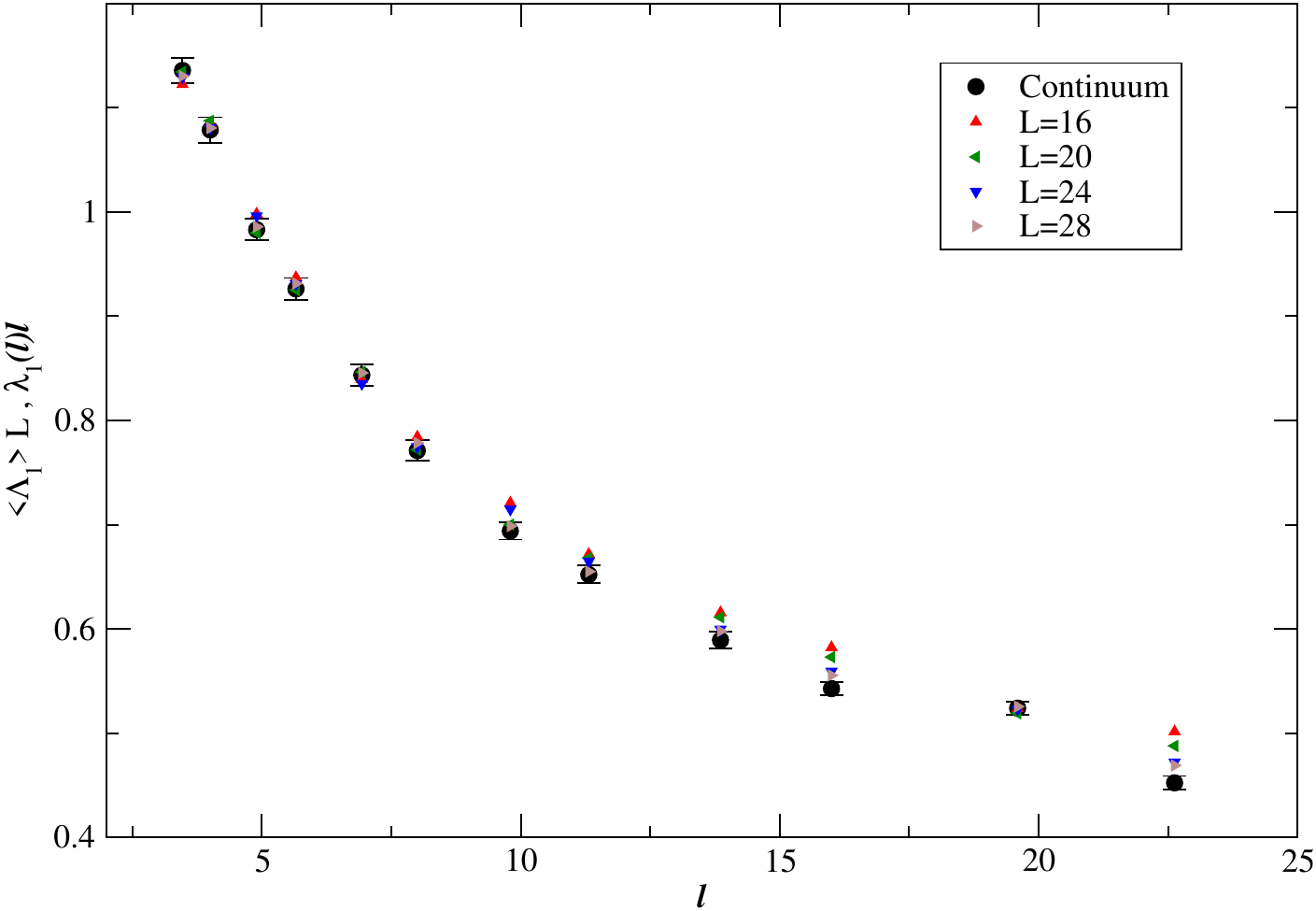}
\includegraphics[scale=0.34]{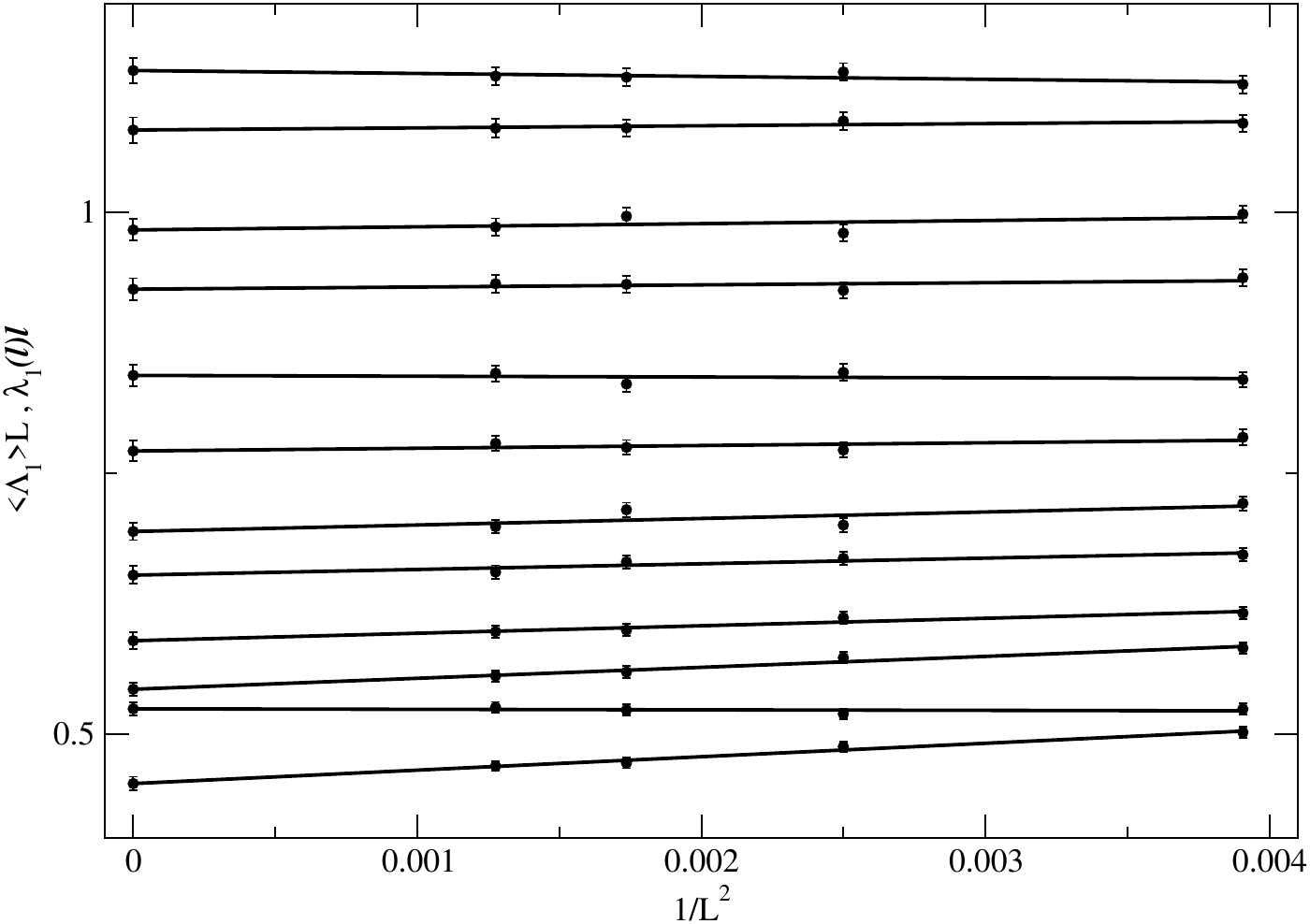}
\includegraphics[scale=0.34]{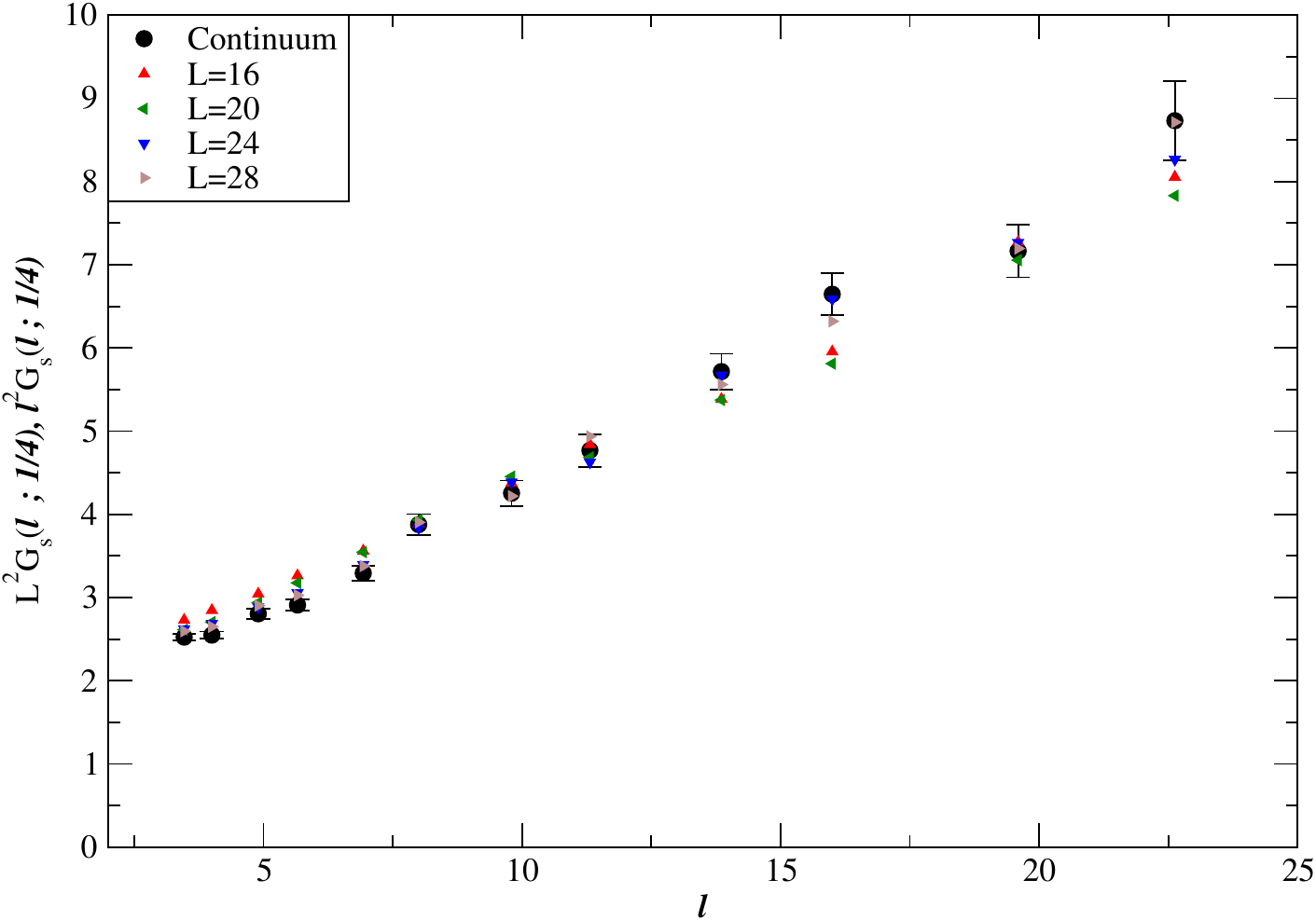}
\includegraphics[scale=0.34]{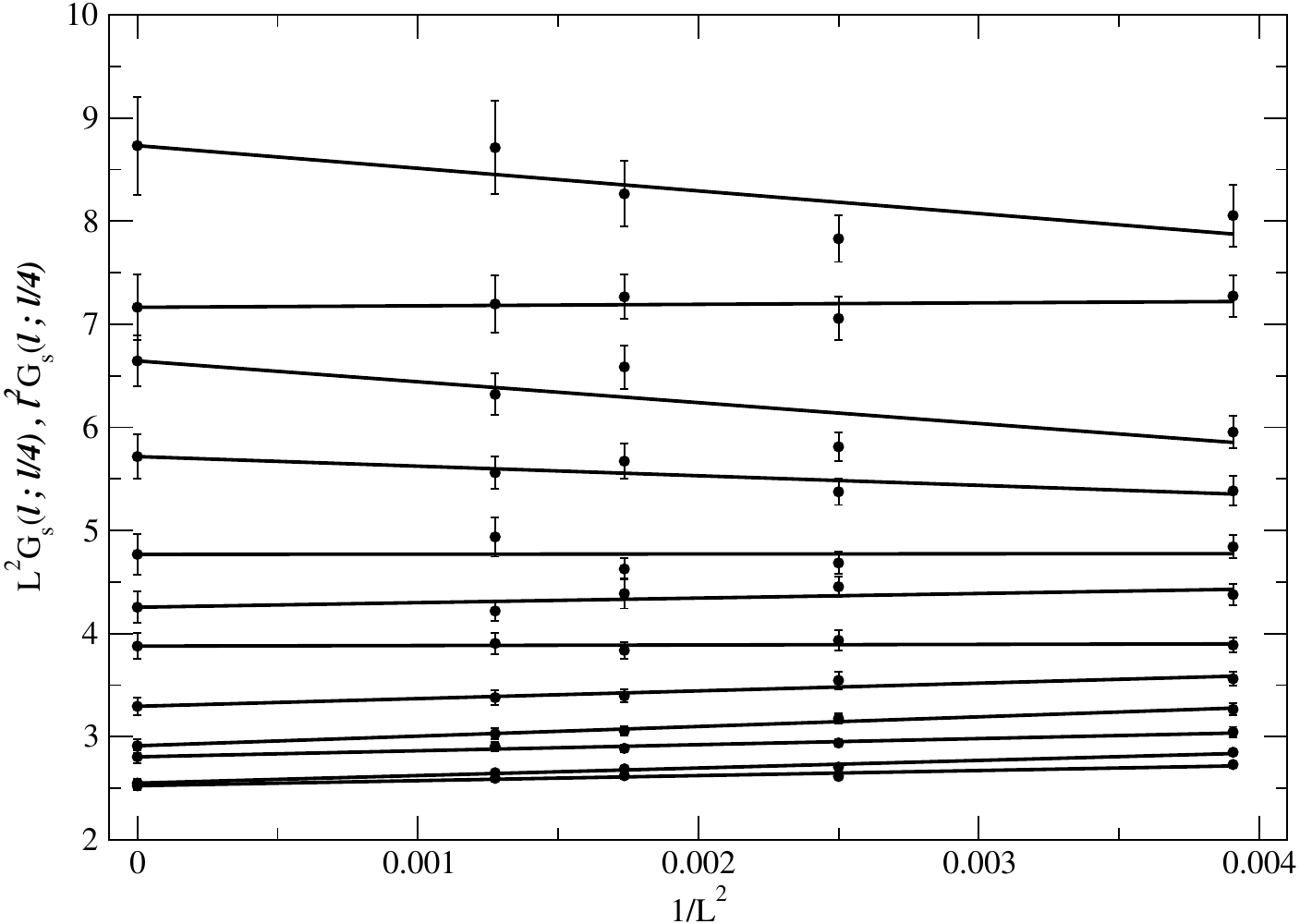}
\caption{ The lowest eigenvalue (top panels) and the scalar correlator (bottom panels) after appropriate $\ell$-scaling are shown as a function of $\ell$ (left panels) and function of $L$ (right panels) for a representative case $(N_c,N_f)=(2,1)$. Only the central values of the data are shown at four values of the lattice extent $L$ in the left panels to avoid clutter, and they are compared against the continuum extrapolated values shown as black-filled circles. The right panels include the errors at each $L$ and show the fit to extract the continuum values.
} \label{fig:Linfext}
\end {figure}

\begin {figure}
\includegraphics[scale=0.46]{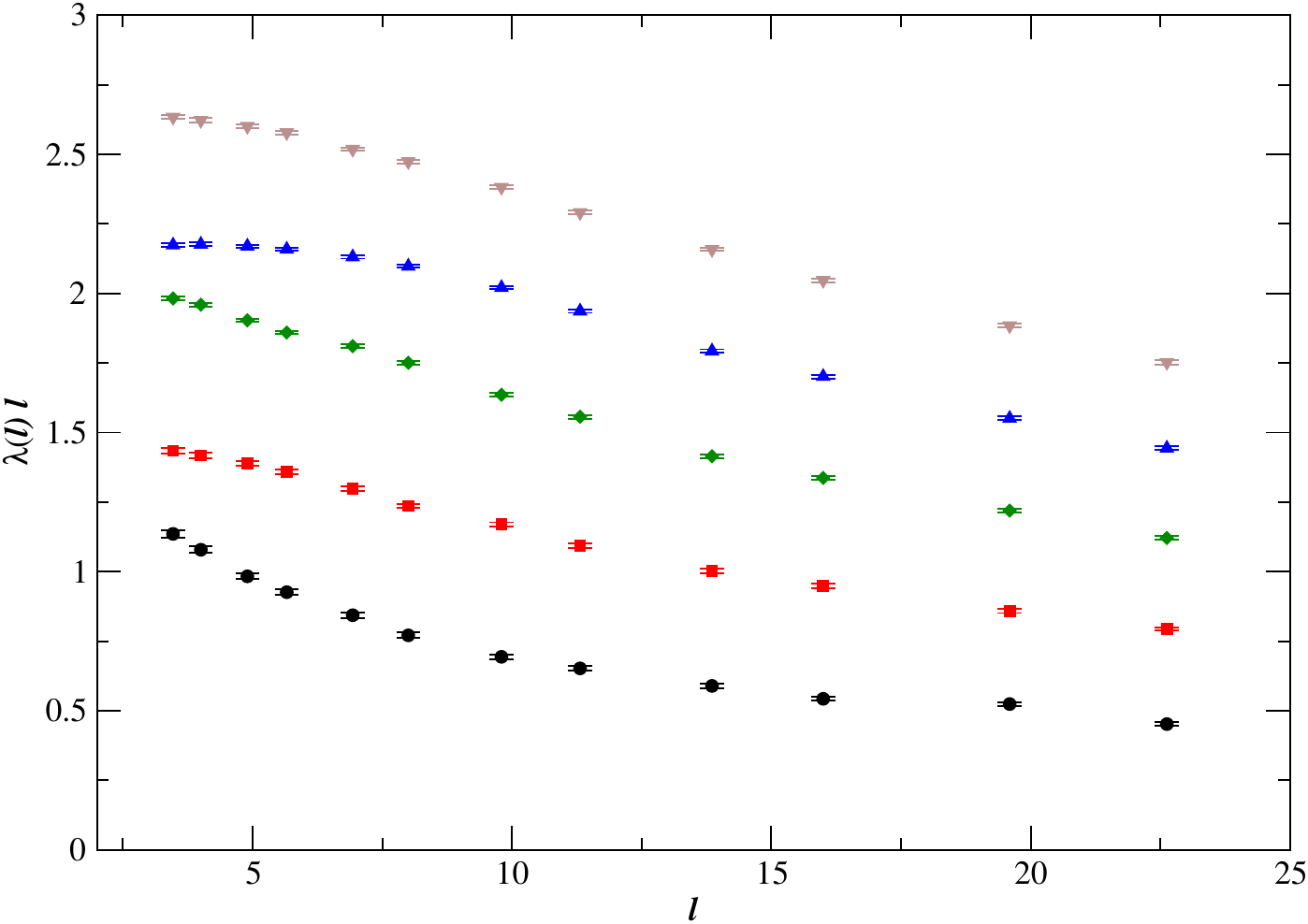}
\includegraphics[scale=0.46]{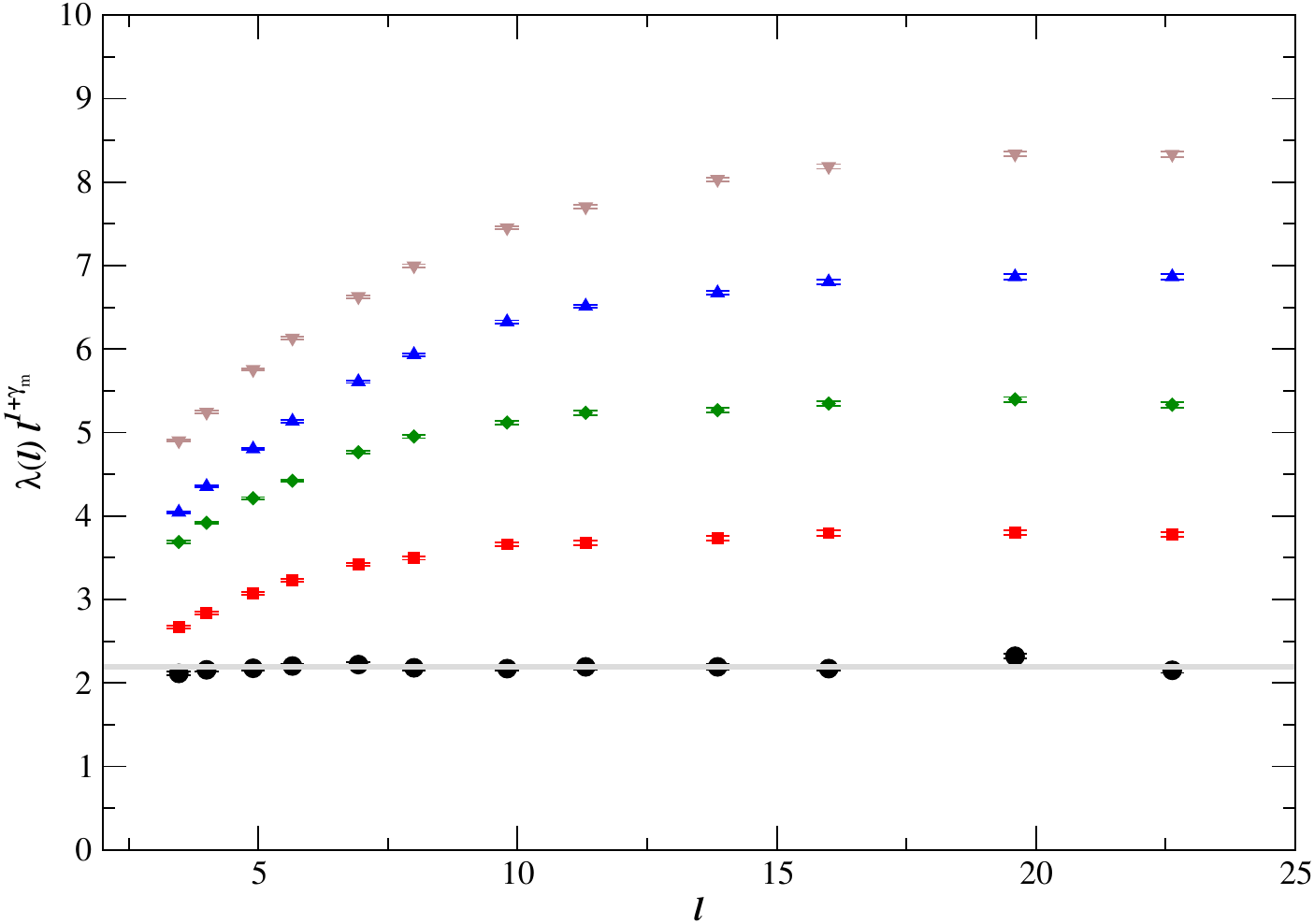}
\caption{The five lowest eigenvalues are plotted as a function of $\ell$ for $N_c=2$ and $N_f=1$. The top panel is scaled as per the naive ultraviolet scaling and the bottom panel is scaled as per the predicted infrared scaling using the appropriate WZW model, namely, $\gamma_m=\frac{1}{2}$.} \label{fig:eigenvalue1}
\end {figure}

\begin {figure}
\includegraphics[scale=0.46]{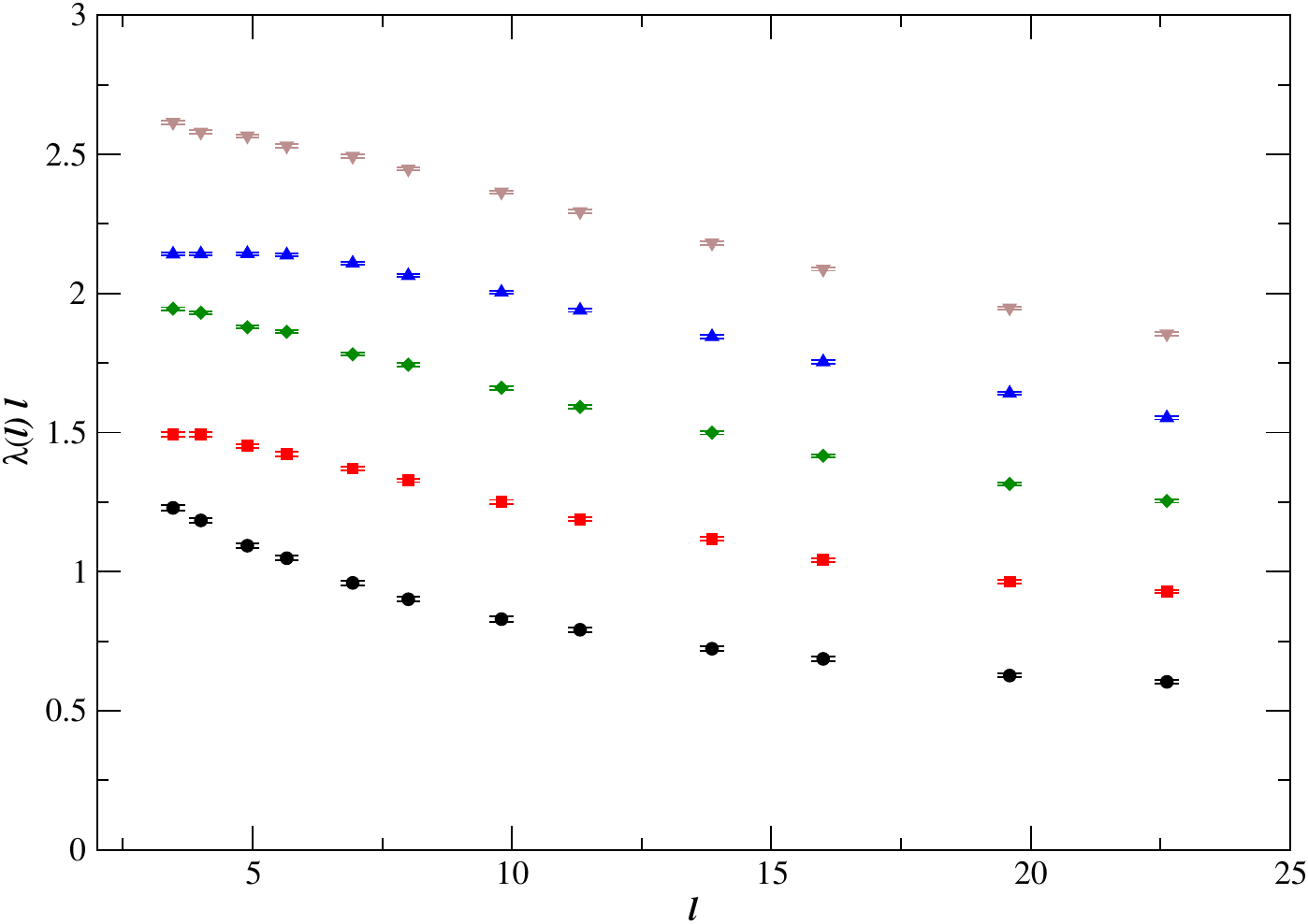}
\includegraphics[scale=0.46]{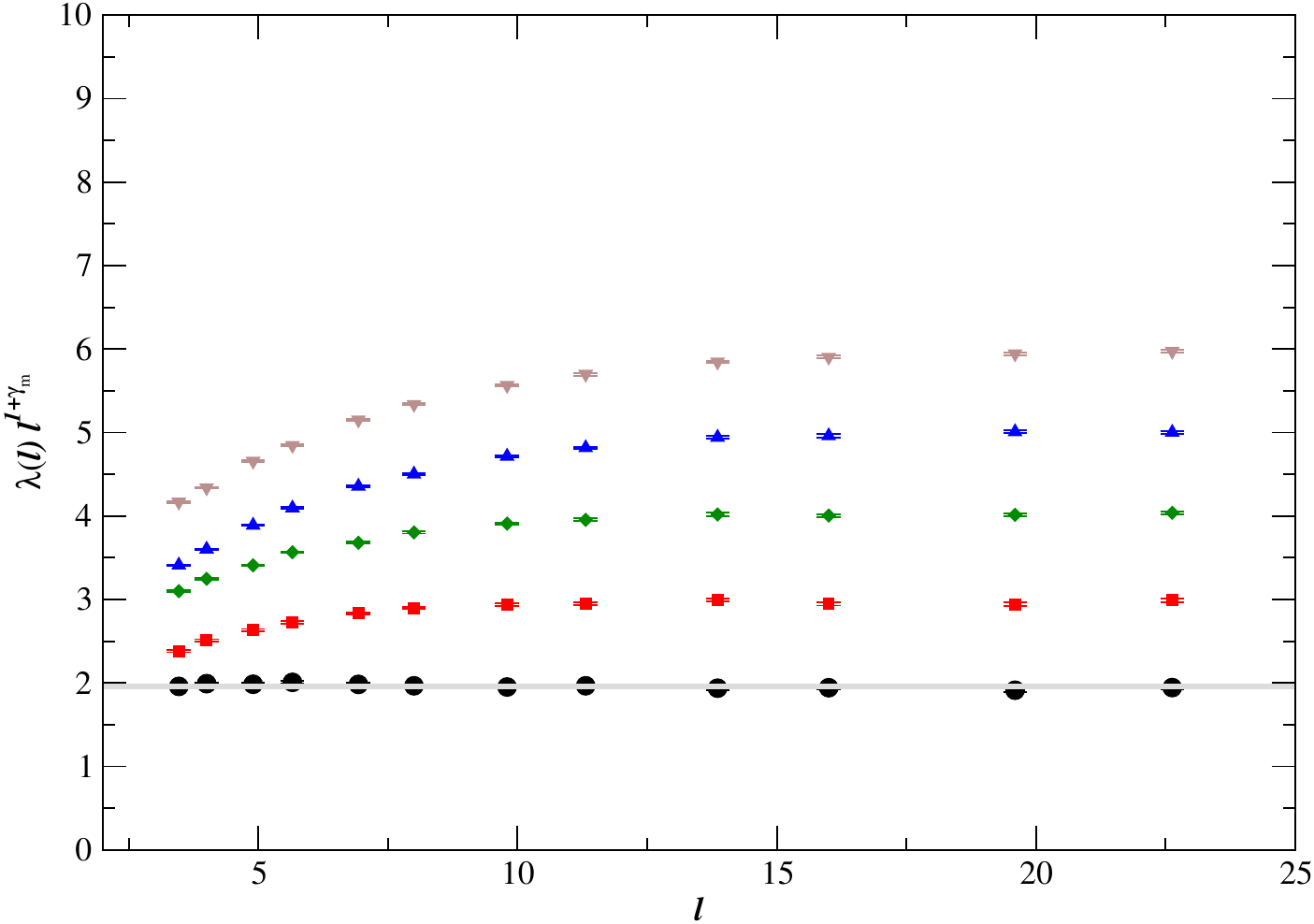}
\caption{The five lowest eigenvalues are plotted as a function of $\ell$ for $N_c=2$ and $N_f=2$. The top panel is scaled as per the naive ultraviolet scaling and the bottom panel is scaled as per the predicted infrared scaling using the appropriate WZW model, namely, $\gamma_m=\frac{3}{8}$.
} \label{fig:eigenvalue2}
\end {figure}
\begin {figure}
\includegraphics[scale=0.46]{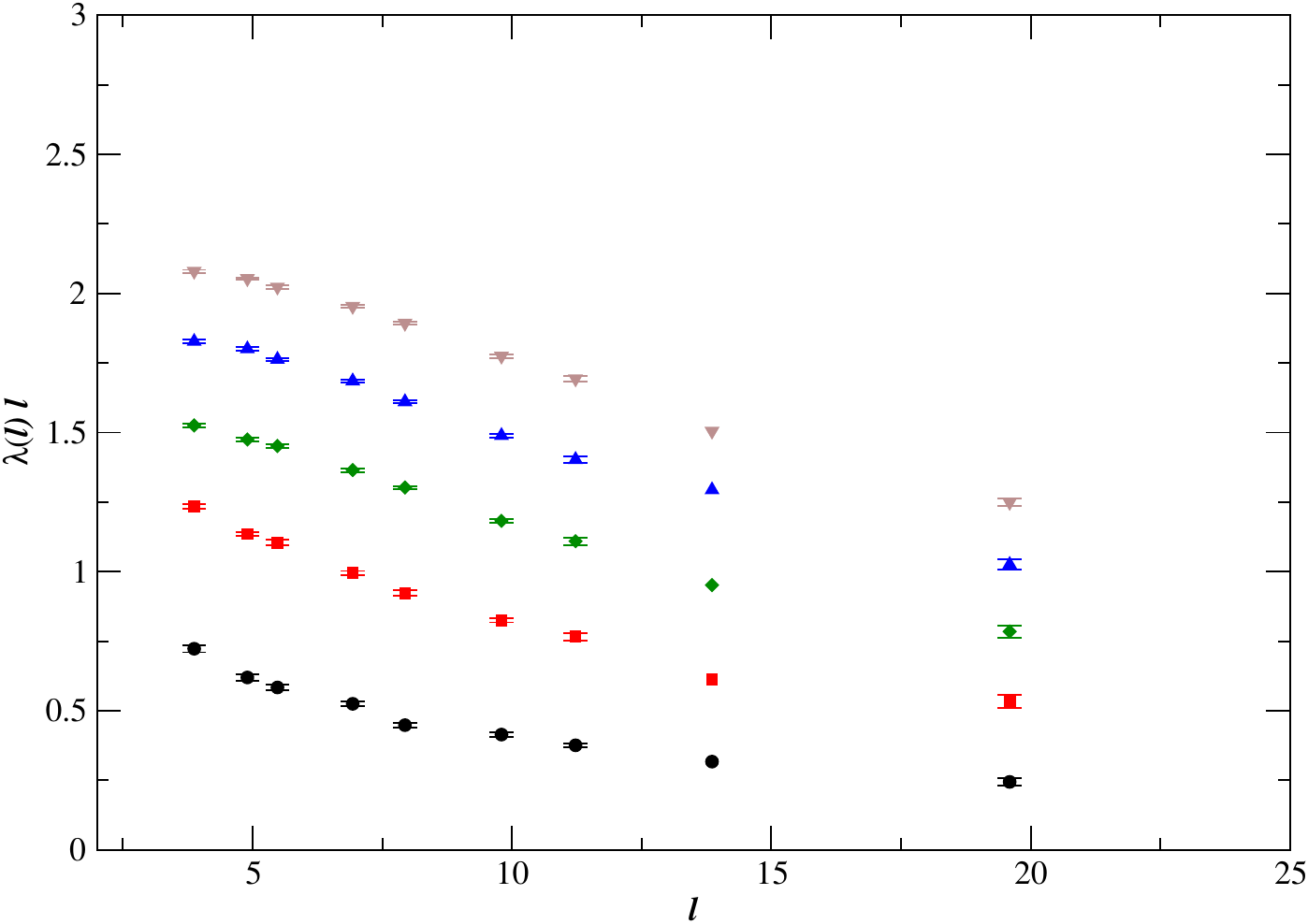}
\includegraphics[scale=0.46]{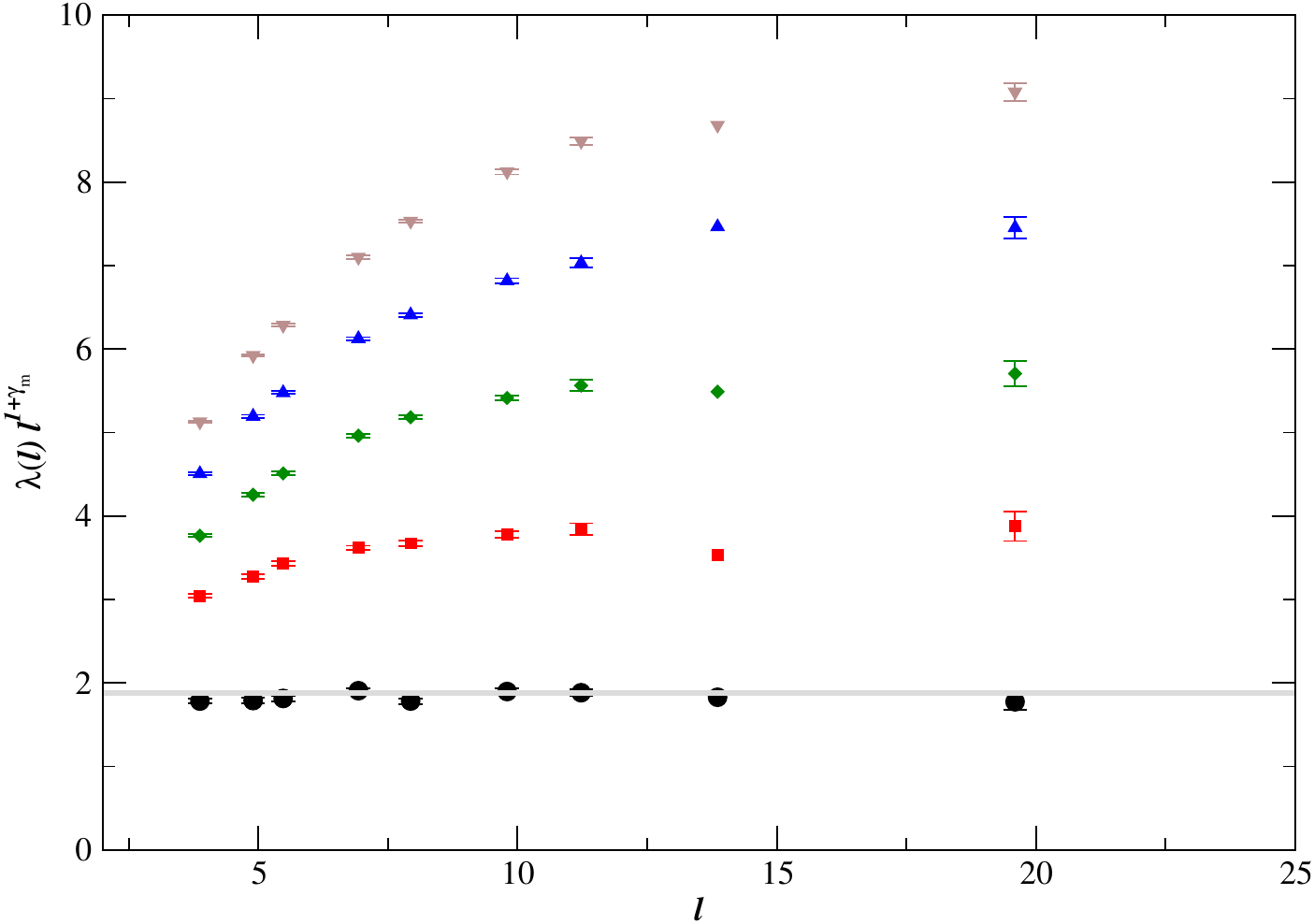}
\caption{The five lowest eigenvalues are plotted as a function of $\ell$ for $N_c=3$ and $N_f=1$. The top panel is scaled as per the naive ultraviolet scaling and the bottom panel is scaled as per the predicted infrared scaling using the appropriate WZW model, namely, $\gamma_m=\frac{2}{3}$.
} \label{fig:eigenvalue3}
\end {figure}
\begin {figure}
\includegraphics[scale=0.46]{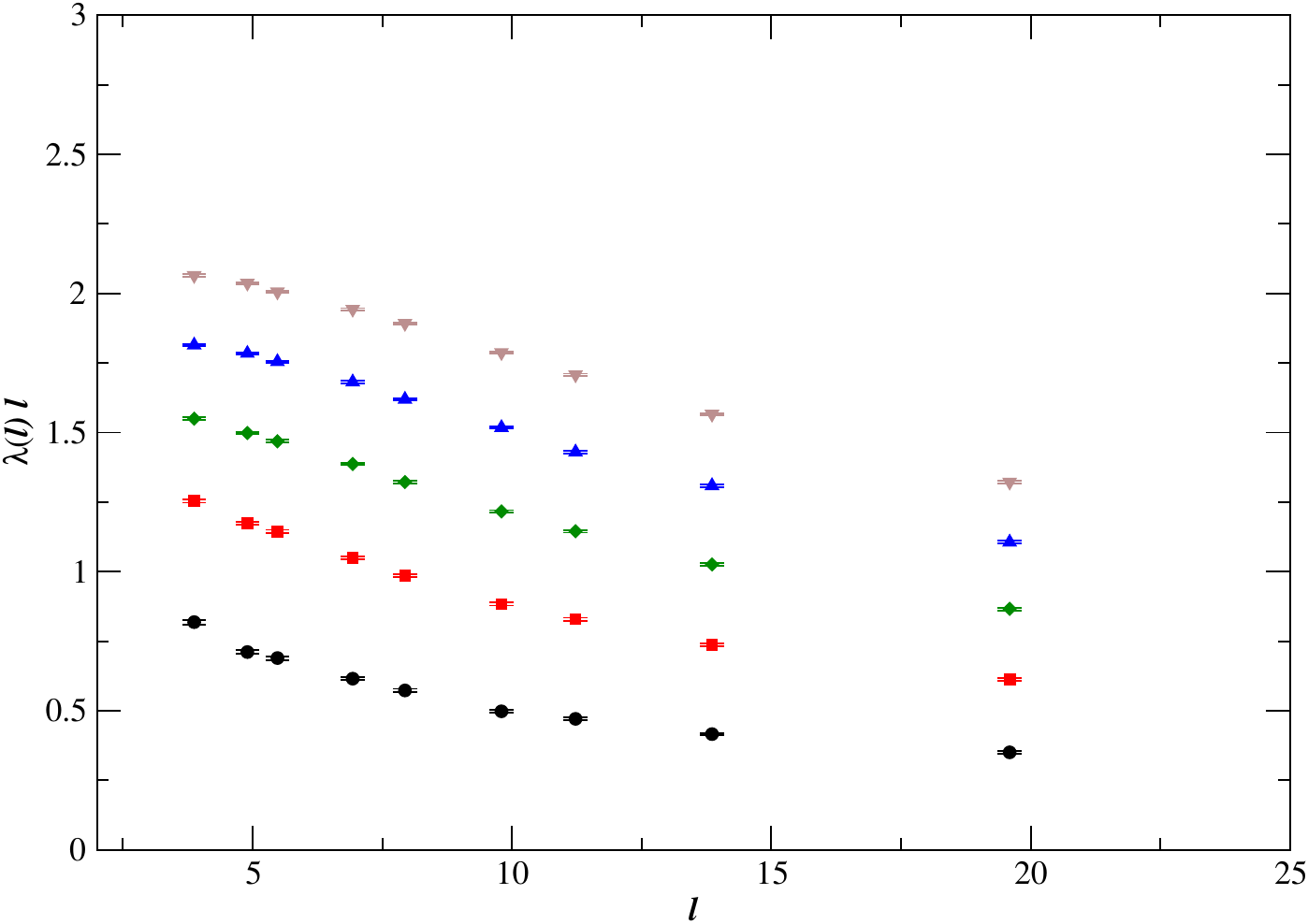}
\includegraphics[scale=0.46]{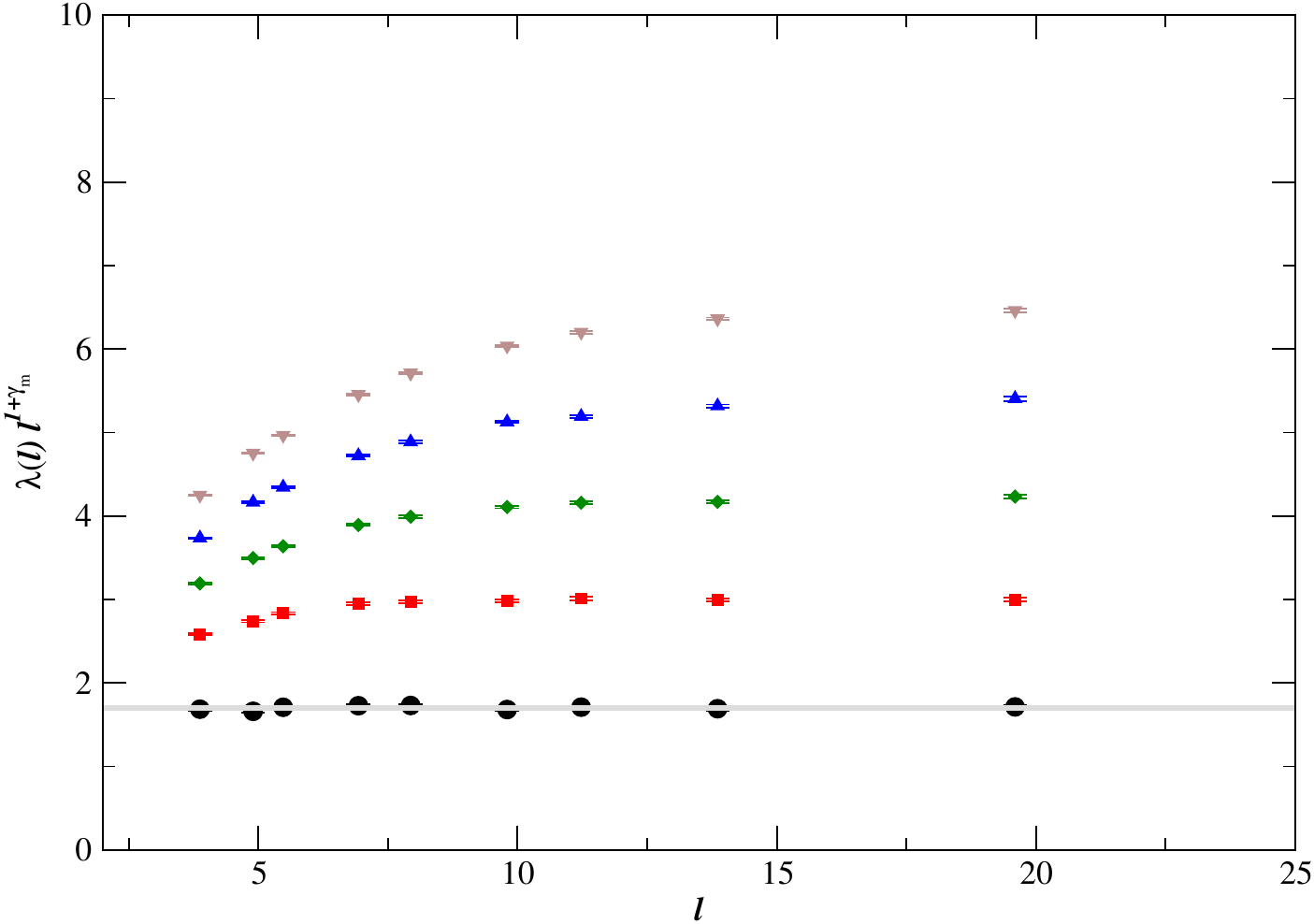}
\caption{The five lowest eigenvalues are plotted as a function of $\ell$ for $N_c=3$ and $N_f=2$. The top panel is scaled as per the naive ultraviolet scaling and the bottom panel is scaled as per the predicted infrared scaling using the appropriate WZW model, namely, $\gamma_m=\frac{8}{15}$.
} \label{fig:eigenvalue4}
\end {figure}

The numerical analysis parallels methods used in the study of three-dimensional massless QED in~\cite{Karthik:2016ppr} and we do not provide the details here.
We used lattices with $L=16,20,24,28$ in our numerical simulations. 
We studied the theories with $(N_c,N_f)=(2,1),(2,2),(3,1),(3,2)$. 
At each $L$, we performed Monte Carlo simulations at the following values of 
dimensionless areas $\ell^2$:
\bea
\ell^2=12,16,24,32,48,64,192,256,384,512 &{\rm for}& N_c=2, \cr
\ell^2=15,24,30,48,63,96,126,192,384 &{\rm for}& N_c=3.
\eea
The values chosen were arbitrary, but cover small-box sizes that are likely to
be closer to ultraviolet behavior, and larger box sizes that are likely to be in the basin of the 
infrared-behavior. At each simulation point, we generated around 5000 trajectories, and we measured the 
eigenvalues and the fermion correlators every 5 trajectories to account for autocorrelations. 
Furthermore, we divided the measurements into 200 jack-knife blocks to estimate the
mean and statistical errors of the observables. We determined the continuum limits of the 
eigenvalues and the two correlators at each fixed $\ell$ by fitting the data from the four values of 
$L$ using a functional form $A+\frac{B}{L^2}$ with fit parameters $A$ and $B$, and extrapolating to $L\to\infty$ limit (i.e., the fitted value of $A$.)
The first lattice correction arises at $O(1/L^2)$ due to the exact chiral symmetry of the lattice overlap fermion. The finite $L$ effects are small for the observables studied in this paper, and \fgn{Linfext} shows the lowest eigenvalue and the scalar correlator at $(N_c,N_f)=(2,1)$ as a sample.

We show the finite-size dependence of the lowest five eigenvalues in \fgn{eigenvalue1} to \fgn{eigenvalue4} 
for $(N_c,N_f)=(2,1),(2,2),(3,1),(3,2)$ cases respectively. As $\ell$ goes from the smallest value to the largest value, we expect a flow from the naive ultraviolet behavior to the infrared behavior predicted by the WZW model in \scn{WZW}. To verify that this is indeed the case, we have plotted $\lambda_i(\ell) \ell$ in the top panels and $\lambda_i(\ell) \ell^{1+\gamma_m}$ in the bottom panels. If the expected ultraviolet and infrared behavior sets in, we should see these appropriately scaled eigenvalues approach a plateau as a function of $\ell$ (for small $\ell$ in the top panels, and larger $\ell$ in the bottom panels). In all four cases studied in this paper, we observe that even the smallest $\ell$ used in the study shows the expected infrared behavior of the lowest eigenvalue, $\lambda_1(\ell)$. To emphasize that $\lambda_1(\ell) \ell^{1+\gamma_m}$ remains a constant throughout the range of $\ell$ considered in this paper, we have drawn a horizontal line through the points to guide the eye. This is clear evidence that $\gamma_m$ for the infrared 
CFT matches that of a WZW model. The second and fourth eigenvalues seem to approach the naive ultraviolet behavior for the smallest $\ell$ studied here. On the other hand, the bottom panels clearly show that $\lambda_i(\ell)$ for $i=2,3,4,5$ flow toward the expected infrared behavior as one approaches large values of $\ell$ and the lower eigenvalues approach the infrared behavior faster than the higher eigenvalues. That the case of $(N_c,N_f)=(3,1)$ has an underlying supersymmetry~\cite{Delmastro:2022prj} does not seem to single out the flow.

\begin {figure}
\includegraphics[scale=0.46]{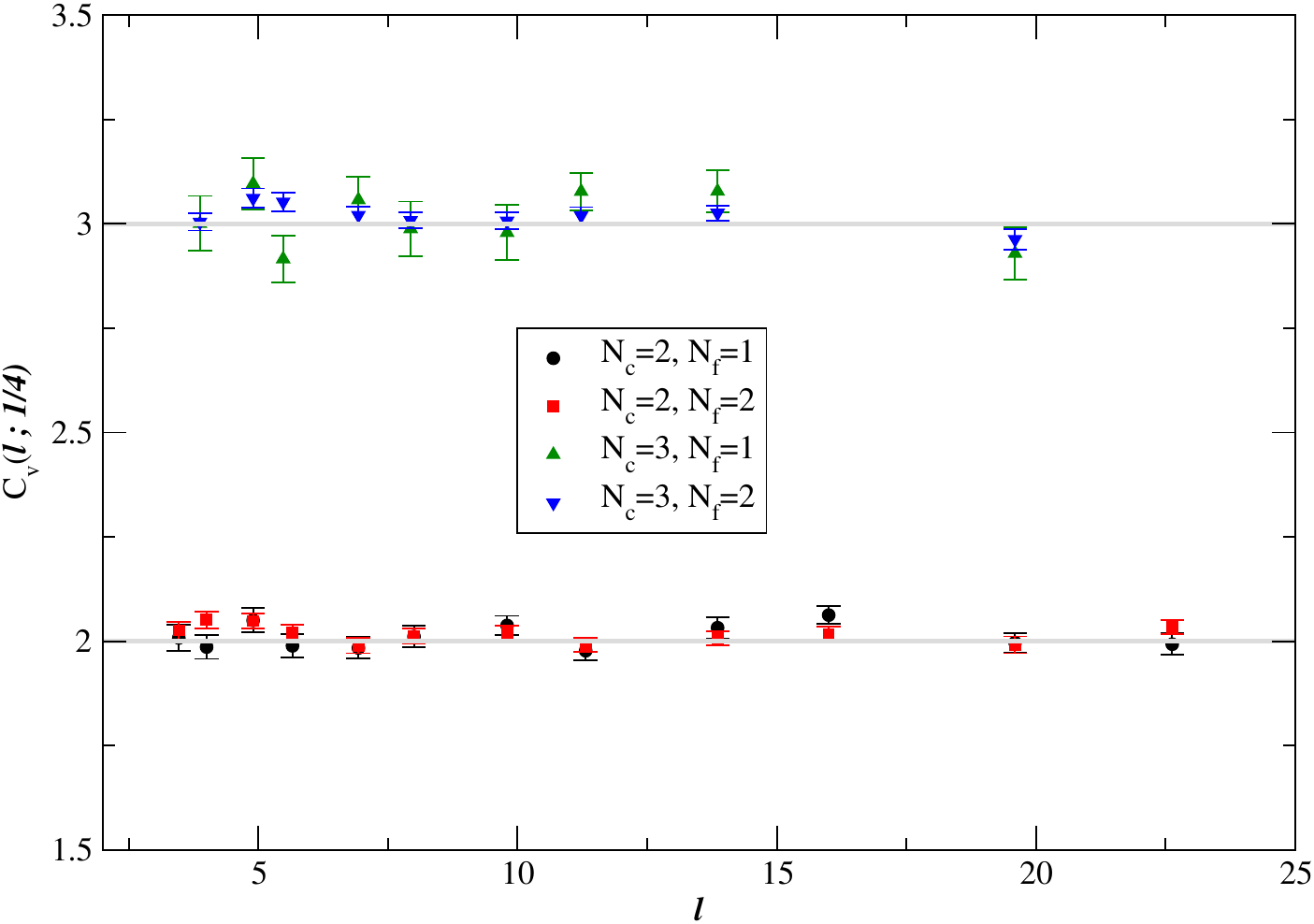}
\caption{ 
The flow of the amplitude of the vector meson correlator, 
as defined in \eqn{cvuvir}. The figure shows the $\ell$-dependence of $C_v(\xi,\ell)$ that should plateau at the UV amplitude for small $\ell$, and should also plateau at the IR amplitude for large-$\ell$. The data from four different $(N_f,N_c)$ are shown using a specific $\xi=1/4$. The horizontal lines are shown to indicate the absence of a UV-IR flow of the vector meson amplitude.
}\label{fig:correlatorv}
\end{figure}
\begin {figure}
\includegraphics[scale=0.46]{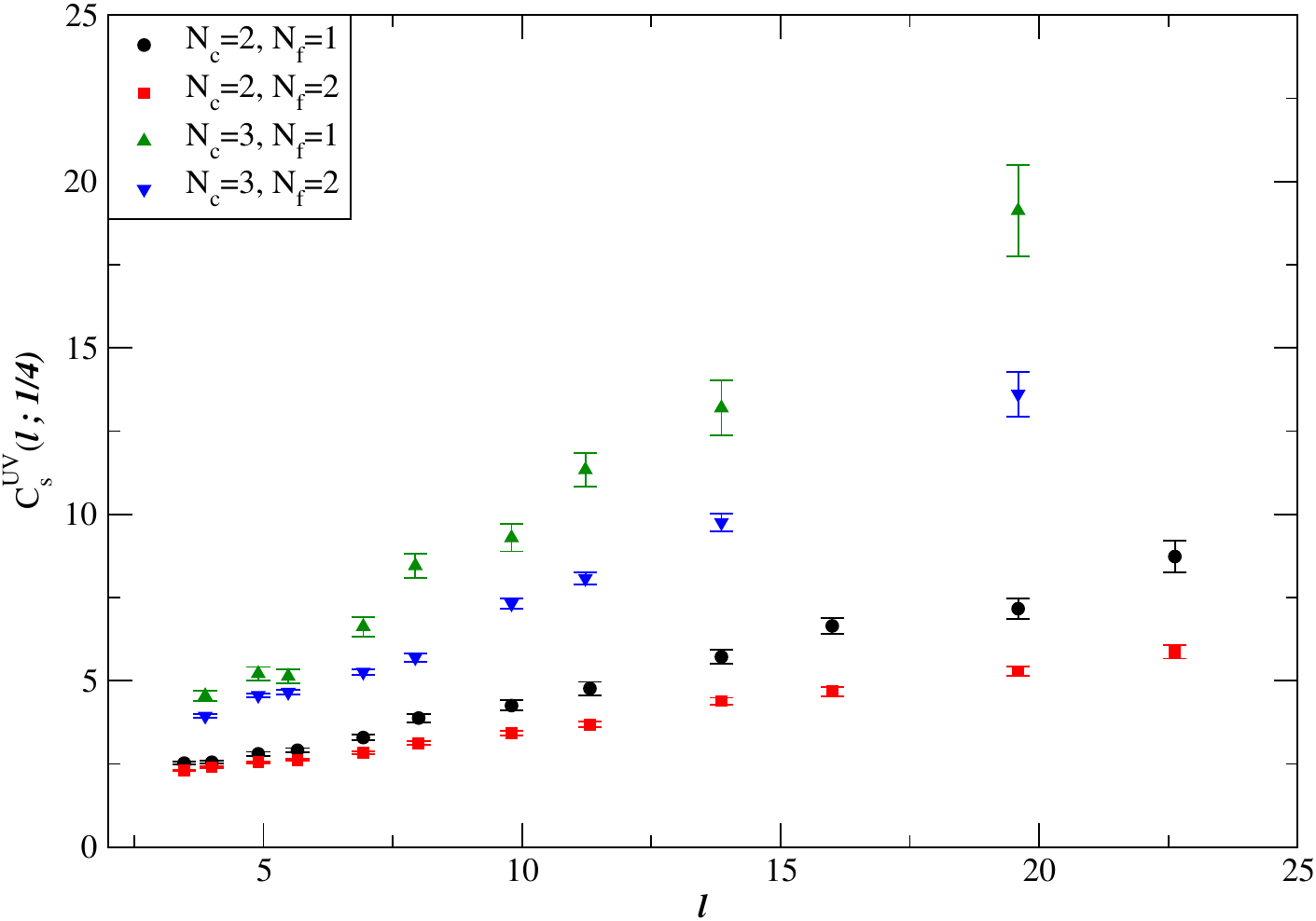}
\includegraphics[scale=0.46]{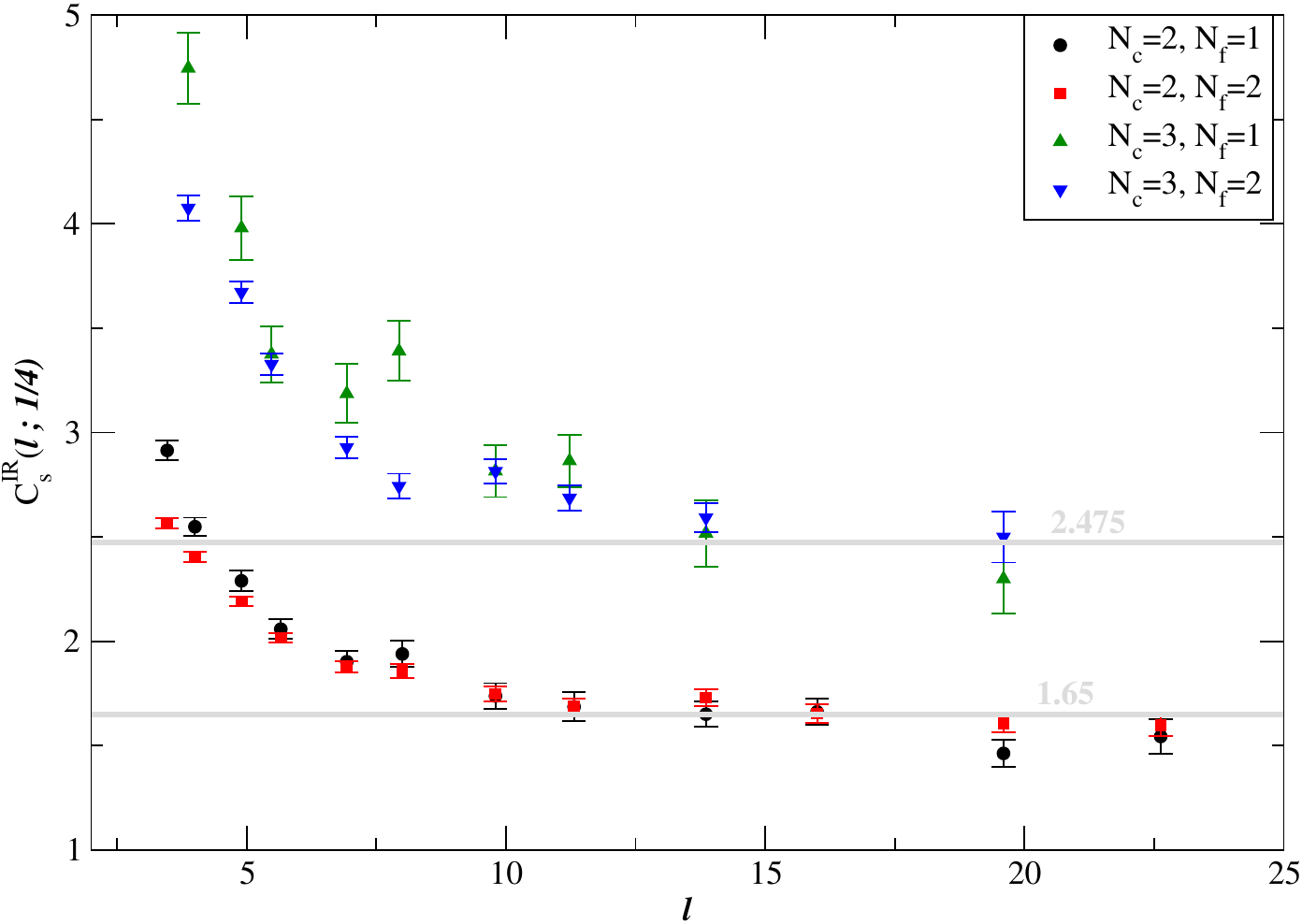}
\caption{The flow of the amplitude of the scalar correlator, 
as defined in \eqn{csuvir} by a UV scaling (top panel) and by WZW infrared scaling (bottom panel) of the scalar correlator. The top panel shows the $\ell$-dependence of $C_s^{\rm UV}(\xi,\ell)$ that should plateau at the UV amplitude for small $\ell$. The bottom panel shows the $\ell$-dependence of $C_s^{\rm IR}(\xi,\ell)$ that should plateau at the IR amplitude for large-$\ell$.  The data from four different $(N_f,N_c)$ are shown using a specific $\xi=1/4$.
}\label{fig:correlators}
\end{figure}

We discuss the flow of the two-point functions below, starting from the conserved vector mesons. 
In \fgn{correlatorv}, we show the continuum extrapolated amplitude, $C_v(\ell,\xi)$ at $\xi=\frac{1}{4}$ computed in the interacting theory. This quantity should flow from its ultraviolet value at small $\ell$ to its infrared value at large $\ell$, with perhaps a crossover between the ultraviolet and infrared behavior at some intermediate $\ell$. However, we note a 
clear plateau over the entire range of $\ell$ we studied with no sign of such a UV-IR crossover. 
This is consistent with the expectation that the vector meson amplitude is a renormalization group 
invariant, namely, $C_v(\ell,\xi=0)$ is independent of $\ell$, owing to the 't Hooft anomaly matching condition for the global flavor symmetry~\cite{Delmastro:2022prj}.
This behavior could be contrasted with the renormalization
group flow of the current amplitudes in three-dimensional
massless QED (see for instance~\cite{Giombi:2016fct,Karthik:2017hol}.)
We also notice that there is no dependence of the data points in \fgn{correlatorv} on $N_f$ and the value of $C_v(\infty, \frac{1}{4})$ is consistent with $N_c$ as expected of the WZW model.

Note that the distinction between $U(N_f)$ and $SU(N_f)$ is not relevant at the level of current correlators in a WZW model. One should note that the free correlator, $G_f(\xi)$ is independent of $N_c$ and $N_f$, and hence, the ratio of 3/2 between the amplitudes is not imposed by hand.

In the following discussion, we return to the scaling dimension of the scalar operator, now discussed using the scalar correlator. 
In \fgn{correlators}, we show the $\ell$-dependence of the $C_s^{\rm UV,IR}(\ell,\xi)$ at a fixed value of $\xi=\frac{1}{4}$.
The ultraviolet behavior is highlighted in the top panel and the infrared behavior is highlighted in the bottom panel.
We should see a plateau in the top panel for small $\ell$ if we are close to the basin of the ultraviolet, free CFT.
Like in the case of the eigenvalues, the smallest value of $\ell$ studied here is still away from the naive ultraviolet limit, as is evident from the residual $\ell$-dependence at small $\ell$ seen in the top panel. Unlike the vector meson correlator, we see a distinct UV-IR flow in the data points in the bottom panel.
The apparent plateau for large values of $\ell$ is again evidence for the approach to a WZW model in the infrared. Like in the case of the vector meson amplitudes, we see that the scalar amplitudes in the infrared are independent of $N_f$ and scale linearly with $N_c$.

\section{Conclusions}
We have studied two-dimensional $su(N_c)$ gauge theories with $N_f$ flavors of massless fermions using the lattice formalism with exact massless fermions on the lattice. We aimed to supplement the analysis usually performed in the Discrete Light-Cone Hamiltonian. In particular, we studied the mass anomalous dimensions and the related scaling dimensions of scalar mesons, which require the presence of both left-chiral and right-chiral conserved flavor currents. Flavor symmetry is exact in our lattice formalism, and our numerical studies with $(N_c,N_f)=(2,1),(2,2),(3,1),(3,2)$ clearly show the expected scaling in the infrared given by the appropriate WZW model. We also see a flow from the ultraviolet to the infrared. We find evidence of the amplitude of the conserved current correlator being renormalization group invariant, as expected from the anomaly matching condition. In contrast, we see a distinct UV-IR flow of the amplitude of the scalar correlator with these amplitudes approaching $N_f$-independent values in the infrared.  Our work closes a conjectural gap in the recent studies on the classification of 
infrared phases of QCD by removing the ultraviolet lattice regulator first before taking the infrared limit.

\acknowledgments
The authors thank Jaume Gomis for useful discussions. R.N. acknowledges partial support by the NSF under grant number
PHY-1913010 and PHY-2310479. S.N. acknowledges support by the Celestial Holography Initiative at the Perimeter Institute for Theoretical Physics and by the Simons Collaboration on Celestial Holography. S.N's research at the Perimeter Institute is supported by the Government of Canada through the Department of Innovation, Science and Industry Canada and by the Province of Ontario through the Ministry of Colleges and Universities. This work used Expanse at SDSC through allocation
PHY220077 from the Advanced Cyberinfrastructure Coordination
Ecosystem: Services \& Support (ACCESS) program, which is supported
by National Science Foundation grants \#2138259, \#2138286, \#2138307,
\#2137603, and \#2138296.
\bibliography{biblio}
\end{document}